
\headline={\ifnum\pageno=1\firstheadline\else
\ifodd\pageno\rightheadline \else\leftheadline\fi\fi}
\def\firstheadline{\hfil}
\def\rightheadline{\hfil}
\def\leftheadline{\hfil}
        \footline={\ifnum\pageno=1\firstfootline\else\otherfootline\fi}
\def\firstfootline{\rm\hss\folio\hss}
\def\otherfootline{\hfil}

\font\tenrm=cmr10

\font\elevenbf=cmbx10 scaled\magstep 1
\font\elevenrm=cmr10 scaled\magstep 1
\font\elevenit=cmti10 scaled\magstep 1

\font\ninerm=cmr9

\nopagenumbers
\hsize=6.0truein
\vsize=8.5truein
\parindent=1.5pc
\baselineskip=10pt
%
\def\ick{\eqalign}
\def\sla{\raise.16ex\hbox{$/$}\kern-.57em}
\def\Dsl{\kern.2em\raise.16ex\hbox{$/$}\kern-.77em\hbox{$D$}}
\def\parsl{\sla\hbox{$\partial$}}
\def\Asl{\kern.2em\raise.16ex\hbox{$/$}\kern-.77em\hbox{$A$}}
\def\bsl{\kern.2em\raise.16ex\hbox{$/$}\kern-.77em\hbox{$b$}}
\def\bsl{\sla\hbox{$b$}}

\def\sla{\raise.16ex\hbox{$/$}\kern-.57em}
\def\Dsl{\kern.2em\raise.16ex\hbox{$/$}\kern-.77em\hbox{$D$}}
\def\parsl{\sla\hbox{$\partial$}}
\def\Asl{\kern.2em\raise.16ex\hbox{$/$}\kern-.77em\hbox{$A$}}
\def\Bsl{\kern.2em\raise.16ex\hbox{$/$}\kern-.77em\hbox{$B$}}
\def\square{\kern1pt\vbox{\hrule height 1.2pt\hbox{\vrule width 1.2pt\hskip 3pt
   \vbox{\vskip 6pt}\hskip 3pt\vrule width 0.6pt}\hrule height 0.6pt}\kern1pt}

\def\ga{\gamma}

\def\mpl{Mod. Phys. Lett. }
\def\np{Nucl. Phys. }

\def\pr{Phys. Rev. }
\def\prd{Phys. Rev.  }

\line{\elevenrm\hfill UdeM-LPN-TH-94-201}
\vglue 5pt
\line{\elevenrm\hfill hep-th/9407029}
\vglue 1.0cm
\centerline{\elevenbf MASS GENERATION FOR GAUGE FIELDS}
\vglue 7pt
\centerline{{\elevenbf WITHOUT SCALARS}
\footnote{\hbox{$^*$}} { \ninerm\baselineskip=11pt
Talk given by D.C. at the MRST94 Conference, ``What Next? Exploring the
Future of High Energy
Particle Physics", McGill University, Montreal, Canada, May 1994.
\hfil} }
\vglue 1.0cm
\centerline{\elevenrm DIDIER CAENEPEEL$^1$ and MARTIN LEBLANC$^{1,2}$}
\baselineskip=13pt
\centerline{\elevenit $^1$ Laboratoire de Physique Nucl\'eaire,}
\baselineskip=12pt
\centerline{\elevenit $^2$ Centre de Recherches Math\'ematiques,}
\baselineskip=12pt
\centerline{\elevenit Universit\'e de Montr\'eal, Montr\'eal, Qc, Canada,
    H3C-3J7}

\vglue 0.8cm
\centerline{\tenrm ABSTRACT}
\vglue 0.3cm
  {\rightskip=3pc
 \leftskip=3pc
 \tenrm\baselineskip=12pt
 \noindent
We present an alternative to the Higgs mechanism to generate masses
for non-abelian gauge fields in (3+1)-dimensions. The initial Lagrangian is
composed of a fermion with current-current and dipole-dipole
type self-interactions
minimally coupled to non-abelian gauge fields. The mass generation occurs
upon the fermionic functional integration. We show that by fine-tuning the
coupling constants the effective theory contains massive non-abelian gauge
fields without any residual scalars or other degrees of freedom.
\vglue 0.8cm }
\line{\elevenbf 1. Introduction \hfil}
\baselineskip=14pt
\elevenrm
\vglue 0.4cm
The standard model has been widely accepted as {\elevenit the} theory of
electro-weak interactions. It has successfully accounted for all experiments
to date, making it perhaps one of the greatest successes of modern
theoretical physics. However, apart from
the unknown value of the top quark mass, one of the present mysteries in the
standard model is the absence of the Higgs particle in present experiments.
This fundamental particle is a scalar, which relegates it as the only one in
this category. Perhaps, we should see these arguments as indications for
looking for alternatives to the Higgs mechanism for generating masses for the
elementary fermions and vector bosons.

Over the last twenty years, other mechanisms to
account for mass generation in the standard model have been proposed such as
technicolor theories ${}^1$, and dynamical symmetry breaking via a
top-quark condensate ${}^2$
in analogy with BCS theory of superconductivity and the
Nambu-Jona-Lasinio mechanism in nuclear structure ${}^3$. The latter
mechanism generates the Higgs particle (not as a fundamental particle)
and its consequences with a four-fermion interaction.

In (2+1)-dimensions, it is well-known
that the addition of a topological Chern-Simons term to the gauge field
kinetic part provides a mechanism for gauge fields mass generation without
spoiling gauge invariance ${}^4$. In a relatively similar spirit,
in (3+1)-dimensions, attempts to reproduce the Chern-Simons term involves
a product of field strength tensors $\epsilon^{\mu\nu\alpha\beta}
F_{\mu\nu}\,F_{\alpha\beta}$. However, since this expression can be written
as a total derivative, it cannot bring any modifications to physics.
It is therefore necessary to introduce another field if we want to
mimic the Chern-Simons mechanism in (3+1)-dimensions. For instance, Freedman
and Townsend (F-T) and others
have developed theories containing an antisymmetric tensor
field coupled to an abelian gauge field ${}^{5,6,7,8}$

Recently, a mechanism for photon mass generation in
(3+1)-dimensions has been suggested ${}^9$,
which consists of a functional integration over fermions
minimally coupled to a low-energy abelian gauge field. The fermions
self-interacts via two types of contact interactions:
current-current and dipole-dipole terms. An antisymmetric tensor
field is introduced via the Hubbard-Stratonovich transformation to perform
the fermion's integration. After imposing conditions on the coupling
constants of the theory, it is possible to write the low-energy effective
action as the abelian model discussed in Ref. [5,6,7,8],
which reproduces a massive abelian gauge field.

Since the massive mediators of forces present in the weak
interactions are known to be of non-abelian nature, we generalize here the
above argument to the case of non-abelian gauge fields ${}^{10}$.
\vglue 0.6cm
\line{\elevenbf 2. Non-Abelian high energy model \hfil}
\vglue 0.4cm
We begin with the following non-renormalizable but SU(N) gauge invariant
Lagrangian at the high-energy scale $\Lambda$
in which a fermion field is minimally
coupled to non-abelian gauge fields
$$
{\cal L}= -{1\over 2} {\rm tr} F_{\mu\nu}F^{\mu\nu}
          + \bar\psi(i\Dsl -m)\psi
          -g_1 {\rm tr} j_\mu j^\mu - g_2 {\rm tr} j_{\mu\nu} j^{\mu\nu}
\eqno (1)
$$
where $D_\mu=\partial_\mu - ig A_\mu$ is the covariant derivative.
The non-abelian gauge fields are
defined by $A_{\mu }=A_{\mu}^a T^a$ where $T^a$ are the Lie algebra
generators obeying the commutation relations
$[T^a,T^b]=iC^{abc}T^c$ and the trace relation ${\rm tr} \Bigl ( T^a T^b\bigr )
={1\over 2} \delta^{ab} $.

The last two quantities in (1) are the current-current
and dipole-dipole fermionic self-interactions. The four-vector current and the
dipole current are given respectively
by $j_\mu~=~j_\mu^a~T^a$ and $j_{\mu\nu}=j_{\mu\nu}^a T^a$ with components
$$\ick {
j_\mu^a &= \bar\psi\gamma_{\mu}T^a\psi\cr
j_{\mu\nu}^a &= \bar\psi\sigma_{\mu\nu}\gamma_5 T^a\psi \cr }
\eqno (2)
$$
both of which transform in the adjoint representation of the SU(N) gauge
group.

In components, the Lagrangian (1) can be written as
$$
\eqalign{
        {\cal L} =&-{1\over 4}F^a_{\mu\nu} F^{a,\mu\nu}
        + \bar\psi(i\Dsl -m)\psi\cr
        &-{g_1\over
2}(\bar\psi\gamma^{\mu}T^a\psi)(\bar\psi\gamma_{\mu}T^a\psi)
        -{g_2\over 2}(\bar\psi\sigma^{\mu\nu}\gamma_5 T^a\psi)
        (\bar\psi\sigma_{\mu\nu}\gamma_5 T^a\psi)\cr}
\eqno (3)
$$
where the field strength is given by
$$
F_{\mu\nu}^a=\partial_{\mu}A_{\nu}^a-\partial_{\nu}A_{\mu}^a +
gC^{abc}A_{\mu}^b A_{\nu}^c.
\eqno (4)
$$
This model is interesting because of the form of the dipole-dipole interaction,
which makes it different from the one studied in Ref. [2].

{}From now on, for definiteness and due to obvious potential applications,
we will consider only the SU(2) gauge group with the usual su(2) Lie
algebra given by the Pauli matrices $T^a={\tau^a\over 2}$, $a=1,2,3$
and structure constants given by $C^{abc}=\epsilon^{abc}$.

We next apply the Hubbard-Stratonovich transformation by
introducing auxiliary non-abelian antisymmetric tensor
fields $b_{\mu\nu}$, which belong
also in the su(2)-Lie algebra, transform according to the adjoint
representation and have mass dimension $[b^a_{\mu\nu}]=1$.
Their introduction permit us
to rewrite the dipole-dipole interaction as
$$
-{g_2\over 2}(\bar\psi\sigma^{\mu\nu}\gamma_5 {\tau^a\over 2} \psi)^2\to
-{1\over 2g_2} b^a_{\mu\nu}b^{a \mu\nu} +
ib_{\mu\nu}^a(\bar\psi\sigma^{\mu\nu}\gamma_5 {\tau^a\over 2}\psi)
\eqno (5)
$$
since one can regain the original Lagrangian by solving the equation of motion
for $b_{\mu\nu}^a$ and substituting the result in the Lagrangian. As noted in
Ref.[9],
we could also transform, in a similar way, the current-current interaction by
introducing other auxiliary gauge fields $a_{\mu}^a$. In what follows, we
choose instead to consider only the introduction of auxiliary antisymmetric
tensor fields and treat perturbatively the remaining four-fermion term.
\vglue 0.6cm
\line{\elevenbf 3. Fermionic functional integration \hfil}
\vglue 0.4cm
We are interested in evaluating the behavior of the theory at low energy for
the
non-abelian gauge fields and antisymmetric tensor fields. We compute the
effective action keeping the gauge and antisymmetric tensor fields external and
integrating out the fermions :
$$\eqalign{
&e^{i\Gamma_{\rm eff}[A^a,b^a]}=\int [{\cal D}\bar\psi][{\cal D}\psi]
        \exp i\int d^4x\cr
        &\Bigl (-{1\over 4}F_{\mu\nu}^a F^{a\mu\nu}
        -{1\over 2g_2}b_{\mu\nu}^a b^{a\mu\nu}
        +\bar\psi(i\parsl -m+g\Asl^a{\tau^a\over 2}+\bsl^a{\tau^a\over 2})\psi
        -{g_1\over 2}(\bar\psi\gamma^{\mu}{\tau^a\over 2}\psi)^2
        \Bigr )\cr
}\eqno (6)
$$
where
$\bsl^a\equiv i \sigma^{\mu\nu}\gamma_5 b_{\mu\nu}^a$

In order to perform the integration over the fermion fields, we expand the
last term in power series and use the vacuum dominance approximation to write
$$
(\bar\psi\ga^\mu{\tau^a\over 2}\psi)^2\to\langle 0|
(\bar\psi\ga^\mu{\tau^a\over 2}
\psi)^2|0\rangle
\simeq|\langle 0|(\bar\psi\ga^\mu{\tau^a\over 2}\psi)|0\rangle|^2
\equiv\langle j^{\mu a}\rangle^2
\eqno (7)
$$
where $\langle j^{\mu a}\rangle$ is the current expectation value.
Using (7) in (6), we obtain the effective action at order $g_1$ given by
$$
e^{i\Gamma_{\rm eff}[A^a,b^a]}=\exp\left\{i\int d^4x\left(-{1\over 4}
        F_{\mu\nu}^aF^{a\mu\nu}-{1\over 2g_2}b_{\mu\nu}^a b^{a\mu\nu}
        -{g_1\over 2}\langle j^a_{\mu}\rangle^2 \right)\right\}
        e^{i\Gamma^{(f)}_0}
\eqno (8)
$$
where
$$
\Gamma_0^{(f)}=\int [{\cal D}\bar\psi][{\cal D}\psi]\exp\Bigl\{i\int d^4x
\bar\psi(i\parsl -m+g\Asl^a{\tau^a\over 2}+\bsl^a{\tau^a\over 2})\psi\Bigr\}
\eqno (9)
$$
is the contribution coming from the integration of the bilinear part in
fermions.
$\Gamma_0^{(f)}$ may be rewritten as
$$
\eqalign{
        \Gamma^{(f)}_0&=-i {\rm Tr}\log (i\parsl-m+g\Asl^a{\tau^a\over 2}
        + \bsl^a{\tau^a\over 2})\cr
        &=\sum_{n=1}^\infty{i\over n}(-1)^n{\rm Tr}\Bigl({1\over i\parsl-m}
        (g\Asl^a+\bsl^a){\tau^a\over 2}\Bigr)^n \cr}
\eqno (10)
$$
and will be evaluated in a derivative expansion up to two derivatives on
fields. In the last equality the trace is taken over spinor indices, group
indices and momenta.

For reasons that will become clear in the course of the discussion, we need to
compute the effective action up to four-point functions (n=4 in (10)).
In part this is guided by
the presence of the SU(2) gauge invariance and also by
noticing that the self-coupling of the non-abelian gauge fields already
contains terms with four fields due to the nonlinearity of the theory.
Competition against those expressions will result in our model.

Regularization is achieved using
a cut-off and Pauli-Villars methods when gauge invariance has to be
preserved. Note that as stated by Faddeev and Slavnov ${}^{11}$,
Pauli-Villars methods may be extended to regularize in a gauge
invariant way in the present theory.

Once the above contributions are evaluated, we compute the expectation value of
the current given by
$\langle j^{a\lambda}\rangle=\delta\Gamma^{(f)}_0/\delta A^a_{\lambda}$
evaluated at $A^a_{\lambda}=0$.
Readers interested in details of the computation are referred to ref.~[10]. We
stress however here two important results coming from the evaluation: 1) a
$b\wedge F$-type term arises from the combination of
$\Gamma_{\rm eff}[A^a,b^b]$ and
$\Gamma_{\rm eff}[A^a,A^b,b^c]$ at order $g_1^0$; 2) the effective current
calculated from the $b\wedge F$-type term gives, when squared,
the kinetic part for the antisymmetric fields.
\vglue 0.6cm
\line{\elevenbf 4. Low energy effective action \hfil}
\vglue 0.4cm
After integrating out the fermions and evaluating the effective current, we
obtain the following low energy effective Lagrangian for the gauge and
antisymmetric tensor fields
$$
\eqalign{
{\cal L}_{\rm eff}=
&-{1\over 4}F_{\mu\nu}^a F^{a\mu\nu}
        +{1\over 12}H_{\mu\nu\rho}^a H^{a\mu\nu\rho}+{g\over 4\sqrt{g_1}}
        \epsilon^{\alpha\beta\mu\nu}b_{\alpha\beta}^a F_{\mu\nu}^a \cr
&+{1\over 2g_2}\left({16\pi^4\over g_1m^2\log(\Lambda^2/m^2)}\right)
        \left({m^2\over 4\pi^2}g_2\log{\Lambda^2\over m^2}-1\right)
b_{\mu\nu}^a
        b^{a\mu\nu} \cr
&-{1\over 3}\left( {\pi^2\over g_1m^2\log(\Lambda^2/m^2)}\right)
        b^{a\nu}_{\rho}[g^{\rho\mu}\partial^2 -4\partial^{\mu}\partial^{\rho}]
        b_{\mu\nu}^a  \cr
&-{1\over 3}\left( {g\pi^2\over g_1m^2\log(\Lambda^2/m^2)}\right)\epsilon^{abc}
        \left\{2A^{a\mu} [2b_{\mu\nu}^b (\partial^{\alpha} b_{\alpha}^{c\nu})
        +(\partial^{\alpha} b_{\mu\nu}^b) b_{\alpha}^{c\nu}]+\dots \right\} \cr
&+\left( {g^2\pi^2\over g_1 m^2\log(\Lambda^2/m^2)} \right)
        \Delta^{abcd} [2A^{a\mu} A^{b\nu} (b_{\nu\alpha}^c b_{\mu}^{d\alpha}
        + b_{\mu\alpha}^c b_{\nu}^{d\alpha})
        -A^{a\mu} A_{\mu}^b b_{\nu\alpha}^c b^{d\nu\alpha}] \cr}
\eqno (11)
$$
where the ellipsis represents contributions to the current
of the same form as the term displayed in the corresponding square bracket but
with shuffled indices and we have redefined the antisymmetric tensor fields as
$$
\sqrt{g_1}{m\over 4\pi^2}\log(\Lambda^2/m^2)\; b_{\mu\nu}^a
\to b_{\mu\nu}^a.\eqno (12)
$$

We now proceed with approximations that will help us to clean up the
expression of Eq.~(11)
in order to recover the claims stated in the abstract.
To eliminate the mass term for the antisymmetric tensor fields (fourth term in
(11)),
we tune the parameters in such a way that the following equation is satisfied
$$
m^2=\Lambda^2 e^{-4\pi^2/m^2g_2}
\eqno (13)
$$
which is consistent with our perturbative analysis; for large
$\Lambda^2$ we have small coupling constant $g_2$.
We can take care of the other unwanted terms (fifth to last term in Eq.~(11))
by assuming a small ratio of coupling constants
$$
{\pi^2\over g_1m^2\log(\Lambda^2/m^2)}\sim {g_2\over g_1}\ll 1.
\eqno (14)
$$
Note, there are other terms arising from squaring the effective
current that we have not displayed here for aesthetical reasons. These terms
may be dropped in the same way as stated above.  However, we need to be a
bit caution in taking care of them since it is necessary to show more
clearly the energy scale of the derivatives expansion approximation made when
we compute the effective action.
Qualitatively, we may estimate the energy scale
of the derivative as $[\partial]\sim [A,b]\sim M_{\rm bosons} =
{g\over\sqrt{g_1}}$ [see bellow]. This qualitative evaluation combined with the
approximation (14) made on the coupling constants and the fact that we consider
a weak theory ($g<1$) give us a sufficient argument for dropping these
undesirable terms.

Finally, we are left with the following low energy gauge invariant
effective Lagrangian
$$
{\cal L}_{\rm eff}=
-{1\over 4}F_{\mu\nu}^a F^{a\mu\nu}
+{1\over 12}H_{\mu\nu\rho}^a H^{a\mu\nu\rho}
+{1\over 4\sqrt{g_1}}g
\epsilon^{\alpha\beta\mu\nu}b_{\alpha\beta}^a F_{\mu\nu}^a
\eqno (15)
$$
valid up to energy $m$, which belongs to a class of non-Abelian B\^{}F-type
model ${}^{12}$ and describes
two transverse and one longitudinal propagating
modes of mass ${g/{\sqrt g_1}}$ which consist of the three degrees of
freedom of massive non-Abelian gauge bosons with mass ${g/{\sqrt g_1}}$ for
each group color indices$^{10}$.
\vglue 0.6cm
\line{\elevenbf 5. Conclusions \hfil}
\vglue 0.4cm
We have succeeded in functionally integrating the four-Fermi theory to end up
with an effective $b\wedge F$-theory in agreement with the model proposed by
Lahiri ${}^{12}$ but different than the one proposed by Freedman and
Townsend  ${}^{5}$.
It is interesting to note that we did not reproduce Freedman-Townsend's model
because
the nonlinearities in the antisymmetric tensor fields are suppressed by
the cutoff $\Lambda$. This is perhaps unfortunate
because the F-T model has higher reducible vector gauge invariance and behaves
properly for renormalization purpose ${}^{13}$. However, it has been shown
to have unitarity problems in tree-level scattering~${}^{14}$.

The degree of freedom count of the theory reveal that the Lagrangian
of Eq.~(15) describes
indeed, for each color, two massive transverse modes
combined with a massive longitudinal mode of the same mass, which is
necessarily interpreted as the third degree of freedom for the
non-abelian gauge fields.  The mechanism described here appears
as a valuable dynamical mass generation process for non-abelian (and abelian)
gauge fields.
\vglue 0.6cm
\line{\elevenbf Acknowledgements \hfil}
\vglue 0.4cm
We gratefully acknowledge D.~London, R.B.~MacKenzie
and M.B.~Paranjape for useful discussions.
This work was supported in part by the Natural Science and
Engineering Research Council of Canada and the Fonds F.C.A.R. du Qu\'ebec.
\vglue 0.6cm
\line{\elevenbf References \hfil}
\vglue 0.4cm
\medskip
\item{1.} S. Weinberg, {\elevenit \prd}{\bf D19}, 1277 (1979);
L. Susskind, {\elevenit \prd} {\bf D20}, 2619 (1979).
\medskip
\item{2.} W.A. Bardeen, C.T. Hill, M. Lindner, {\elevenit \prd} {\bf D41},
 1647 (1990).
\medskip
\item{3.} Y. Nambu and G. Jona-Lasinio, {\elevenit \pr} {\bf 122}, 345 (1961).
\medskip
\item{4.} R. Jackiw and S. Templeton, {\elevenit \prd} {\bf D23}, 2291 (1981);
J. Schonfield, {\elevenit \np}{\bf B185}, 157 (1981); S. Deser, R. Jackiw and
S. Templeton, {\elevenit Ann. Phys.}{\bf 140}, 372 (1982).
\medskip
\item{5.} D.Z. Freedman and P.K. Townsend, {\elevenit \np}{\bf B177}, 282
(1981).
\medskip
\item{6.} M. Kalb and P. Ramond, {\elevenit \prd}{\bf D9}, 2273 (1974).
\medskip
\item{7.} E. Cremmer and J. Scherk, {\elevenit \np}{\bf B72}, 117 (1974).
\medskip
\item{8.} C.R. Hagen, {\elevenit \prd}{\bf D19}, 2367 (1979).
\medskip
\item{9.} M. Leblanc, R.B. MacKenzie, P.K. Panigrahi, R. Ray,
{\elevenit Int. J. Mod. Phys.}{\bf A9}, (1994), (in press).
\medskip
\item{10.} D. Caenepeel and M. Leblanc, preprint UdeM-LPN-TH-94-194, CRM-2169,
(1994).
\medskip
\item{11.} L.D. Fadeev and A.A. Slavnov, {\elevenit Gauge Fields:
Introduction to Quantum Theory} (Benjamin/Cummings, Massachusetts, 1980),
p.136.
\medskip
\item{12.} A. Lahiri, Los Alamos preprint LA-UT-92-3477 (1992).
\medskip
\item{13.} M. Leblanc, D.G.C McKeon, A. Rebhan, T.N. Sherry, {\elevenit \mpl}
{\bf A6}, 3359 (1991).
\medskip
\item{14.} D.G.C. McKeon, {\elevenit Can. J. Phys.}{\bf 69}, 1249 (1991).

\bye